\documentclass[fleqn,twocolumn,epjc3]{svjour3}
\pdfoutput=1
\usepackage[english]{babel}
\usepackage[utf8]{inputenc}
\usepackage{graphicx} 
\usepackage{float}
\usepackage{amsmath}
\usepackage{mathtools}
\usepackage{anysize}
\usepackage{array}
\usepackage{subfigure}
\usepackage{cite}
\usepackage{appendix}
\usepackage{hyperref}
\usepackage{xcolor}
\usepackage{url}
\usepackage{pdfpages}
\usepackage{lscape}
\usepackage{multirow}
\usepackage{atlasphysics}
\journalname{IFIC/15-94} 
\begin{document}


\title{Constraints on four-fermion interactions from the $t\bar{t}$ charge asymmetry at hadron colliders}
\author{M.~Perelló~Roselló\thanksref{e1,addr1}, M.~Vos\thanksref{e2,addr1}}
\thankstext{e1}{martin.perello@ific.uv.es}
\thankstext{e2}{marcel.vos@ific.uv.es}
\institute{IFIC (UVEG/CSIC), Apartado de Correos 22085, E-46071 Valencia, Spain\label{addr1}}
\date{December 23rd, 2015}

\maketitle

\begin{abstract}
The charge asymmetry in top quark production at hadron colliders is sensitive
to beyond-the-Standard-Model four-fermion interactions. 
In this study we compare the sensitivity of cross-section and charge 
asymmetry measurements to effective operators describing four-fermion 
interactions and study the limits on the validity of this approach. 
A fit to a combination of Tevatron and LHC measurements yields stringent
limits on the linear combinations $C_1$ and $C_2$ of the four-fermion
effective operators. 
\keywords{Top quark, hadron collider, charge asymmetry}
\end{abstract}

\section{Introduction}

Since the discovery of the top quark, its properties and interactions 
have been characterized in some detail. 
The LHC run I analyses are extending the programme initiated at the
Tevatron in several ways. 

All measurements so far are in good agreement with the Standard Model
predictions. The most notorious exception is the measurement of the 
forward-backward asymmetry in $p\bar{p}$ collisions at 1.96~\tev{}
at the Tevatron~\cite{Abazov:2007ab,Aaltonen:2008hc} and its dependence on 
the kinematics of the \ttbar{} system~\cite{Abazov:2015fna,Abazov:2014cca,Aaltonen:2012it,Abazov:2011rq,Aaltonen:2011kc}. 
Excitement has decreased considerably in recent years, as the discrepancy
failed to grow as additional Tevatron data were added.
Taking into account the EW correction~\cite{Hollik:2011ps} and the full 
next-to-next-to-leading-order (NNLO) QCD corrections~\cite{Czakon:2014xsa}
the remaining tension of the inclusive measurements at the Tevatron with
the SM prediction is down to the 1.5 $\sigma$ level.
Measurements of a related charge asymmetry in 7~\tev~\cite{Chatrchyan:2014yta,Aad:2013cea,Chatrchyan:2012cxa,ATLAS:2012an} 
and 8~\tev~\cite{Khachatryan:2015oga,Khachatryan:2015mna,ATLAS-TOPQ-2014-16} $pp$ collisions
at the LHC by ATLAS and CMS are consistent with the SM prediction. 

We assume in the following that all data on the top quark, including the 
Tevatron $A_{FB}$ puzzle, is in reasonable agreement with the SM description.
Remains the task of deriving the most comprehensive constraints on extensions
of the Standard Model. The large number of related measurements requires a 
sophisticated multi-parameter treatment. The effective-operator paradigm
seems an adequate solution to recast the wealth of measurements into
a manageable number of constraints. First steps in the direction
of a global fit to the top sector were recently set by the TopFitter 
collaboration~\cite{Buckley:2015lku,Buckley:2015nca}.

In this paper we derive constraints on
four-fermion operators from measurements at hadron colliders. 
We compare the sensitivity of available and future cross-section and
charge asymmetry measurements, signalling 
the complementarity of both types of measurements. 
We study the limits to the validity of the effective operator 
approach for a number of measurements and propose a practical solution 
to guarantee valid results with the current data and in the foreseeable
future. Finally, we derive constraints on the four-fermion operators
from Tevatron and LHC data and present the prospects for an addition
of future measurements.

\section{Effective operator setup}

A general effective Lagrangian expands around the Standard Model
in terms of $\Lambda^{-2}$:
\begin{equation}
\mathcal{L}_{eff} = \mathcal{L}_{SM} + \frac{1}{\Lambda^2} \sum_i C_i O_i + \mathcal{O}\left(\Lambda^{-4} \right),
\end{equation}
where the scale of new physics $\Lambda$ has to
be taken to several \tev{} for the effective operator paradigm to hold.
We limit our analysis to contributions proportional to ($\Lambda^{-2}$)
(i.e. the interference of the Standard Model with dimension-6 operators). 
In Section~\ref{sec:val} we do, however, estimate the size of the
($\Lambda^{-4}$) terms by calculating the contribution of the
square of the dimension-6 operators. 

In reference \cite{Grzadkowski:2010es} a basis is given for a complete set of 
dimension-six operators. As we are interested in the four-fermion 
operators involved in $t\bar{t}$ production at the LHC, operators 
including leptonic initial states are not included. The reduced group 
of seven four-fermion operators is listed 
in Table \ref{t:4fo}. Operators with the form 
$\left(\bar{q}\lambda^A u^i \right)\left(\bar{u}^j \lambda^A q \right)$ 
can be turned into a linear combination of $O_{qu}^{(1)}$ and are not included. 
These seven operators can be reduced to four by using a flavour-specific 
linear combination \cite{Zhang:2010dr}:

\begin{align}
C_u^1 &= C_{qq}^{(8,1)}+C_{qq}^{(8,3)}+C_{ut}^{(8)} \nonumber \\
C_u^2 &= C_{qu}^{(1)}+C_{qt}^{(1)} \nonumber\\
C_d^1 &= C_{qq}^{(8,1)}-C_{qq}^{(8,3)}+C_{dt}^{(8)} \nonumber\\
C_d^2 &= C_{qd}^{(1)}+C_{qt}^{(1)} 
\label{eq:comb_lin}
\end{align}

A further reduction of the basis for four-fermion operators 
to two effective operators is achieved by assuming 
$C_u^1=C_d^1=C^1$ and $C_u^2=C_d^2=C^2$. This reduction is valid in models 
where the new massive states couple to $u-$type and $d-$type quarks 
with the same strength. Among the models that satisfy this requirement the 
axigluon~\cite{Antunano:2007da} has received most attention in the 
context of the \ttbar{} charge asymmetry measurements at the Tevatron 
and the LHC. We note that the assumption is also valid for models
that are not strictly flavour-universal, such the axigluon 
with an opposite-sign coupling top quarks ($g_t = - g_q$), 
that can give rise to positive contributions to the asymmetry,
and the Kaluza Klein gluon as realized in Randall-Sundrum warped 
extra-dimensions in References~\cite{Agashe:2006hk,Lillie:2007yh},
the main benchmark for direct searches for resonant signals in
\ttbar{} production. 

\begin{table}[h!]
\centering
\begin{tabular}{c}
\hline
Operator \\ \hline 
$O_{qq}^{(8,1)} = \frac{1}{4} \left(\bar{q}^i \gamma_\mu \lambda^A q^j \right) \left(\bar{q} \gamma^\mu \lambda^A q \right)$\\ 
$O_{qq}^{(8,3)} = \frac{1}{4} \left(\bar{q}^i \gamma_\mu \tau^I \lambda^A q^j \right) \left(\bar{q} \gamma^\mu \tau^I \lambda^A q \right)$\\ 
$O_{ut}^{(8)} = \frac{1}{4} \left(\bar{u}^i \gamma_\mu \lambda^A u^j \right) \left(\bar{t} \gamma^\mu \lambda^A t \right)$\\ 
$O_{dt}^{(8)} = \frac{1}{4} \left(\bar{d}^i \gamma_\mu \lambda^A d^j \right) \left(\bar{t} \gamma^\mu \lambda^A t \right)$\\ 
$O_{qu}^{(1)} = \left(\bar{q} u^i \right) \left(\bar{u}^j q\right)$ \\ 
$O_{qd}^{(1)} = \left(\bar{q} d^i \right) \left(\bar{d}^j q\right)$\\ 
$O_{qt}^{(1)} = \left(\bar{q}^i t \right) \left(\bar{t} q^j\right)$ \\ \hline
\end{tabular}
  \caption{Four-fermion operators involved in $t\bar{t}$ production at hadron colliders in the notation from \cite{Zhang:2010dr} where $q$ is the left-handed quark doublet, $u$ and $d$ corresponds to the up and down right-handed quarks of the first two families respectively, and $t$ represents the right-handed top quark. Superscripts $i$, $j$ are used to denote the first two generations. }
  \label{t:4fo}
\end{table}

\section{Measurements}

To constrain the four-fermion effective operator coefficients simultaneously 
we need at least four independent measurements with good sensitivity to these 
operators. We choose the inclusive forward-backward asymmetry measured at 
Tevatron, and the charge asymmetry measured at the LHC at $\sqrt{s} = 8$ TeV. 
The inclusive $t \bar{t}$ production cross-section at the Tevatron and at 
the LHC at $\sqrt{s} = 8$ TeV are also included. The datasets are summarized in 
Table~\ref{t:datasets}. 

\begin{table*}[t]
\centering
\begin{tabular}{lcc}
\hline
 & SM prediction &  Measurement \\ \hline
Tevatron, 1.96~\tev{} $p\bar{p}$, CDF+D0, x-section & $7.16 \pm 0.26$ pb \cite{Czakon:2013goa} & $7.60 \pm 0.41$  pb \cite{Aaltonen:2013wca}\\
Tevatron, 1.96 1.96~\tev{} $p\bar{p}$, CDF+D0, $A_{FB}$ &  $9.5 \pm 0.7 \%$ \cite{Czakon:2014xsa}&  $13 \pm 2.3 \%$ \cite{Abazov:2014cca, Aaltonen:2012it}\\
LHC, 8~\tev{} $pp$, CMS+ATLAS  inclusive $\sigma$ &$245.80 \pm 10.56 $  pb \cite{Czakon:2013goa}  & $241.50 \pm 8.54 $ pb \cite{CMS:2014gta} \\ 
ATLAS 8~\tev{} $pp$, inclusive $A_{C}$ & $1.11 \pm 0.04 \%$ \cite{Bernreuther:2012sx} & $0.9 \pm 0.5 \%$ \cite{Aad:2015noh} \\ 
CMS 8~\tev{} $pp$, inclusive $A_{C}$ & $1.11 \pm 0.04 \%$ \cite{Bernreuther:2012sx} & $0.3 \pm 0.4 \%$ \cite{Khachatryan:2015mna} \\ 
ATLAS 8~\tev{} $pp$, differential $A_{C}$ ($m_{\ttbar{}} >$ 0.75~\tev) & $1.60 \pm 0.04 \%$\cite{Kuhn:2011ri}& $4.2 \pm 3.2 \%$ \cite{Aad:2015lgx}\\ \hline
\end{tabular}
  \caption{Datasets used in the fit. The Tevatron $A_{FB}$ measurement corresponds to a naive approximation between D0 and CDF experiments \cite{Aguilar-Saavedra:2014kpa}. A combination of the ATLAS and CMS measurements of the inclusive asymmetry at 8 TeV is not yet available, so both measurement are kept as independent constraints.}
  \label{t:datasets}
\end{table*}

The selection of Table~\ref{t:datasets} emphasizes inclusive measurements
that integrate over all kinematic regimes. The use of differential 
measurements, especially of the production of high-mass \ttbar{} pairs,
may offer greater sensitivity to high-scale physics beyond the
SM~\cite{Hewett:2011wz}. We therefore include a recent ATLAS result 
for the charge asymmetry in events
where the top quark pair is produced with a large invariant mass~\cite{Aad:2015lgx}, 
which we take as a proxy for measurements in boosted top quark pair production
that become available at the LHC.

\section{Sensitivity to effective operators}
\label{sec:sensitivity}

We generate $t\bar{t}$ samples at parton-level with the 
Monte Carlo generator Madgraph~\cite{Degrande:2011ua, Alwall:2011uj} using 
the UFO model TopEffTh to calculate the impact of the effective 
operators~\cite{Degrande:2010kt} on the cross-section and charge asymmetry.

The dependence of the top quark pair production cross section
and the charge asymmetry on the four-fermion operator coefficients is
parameterized\footnote{As we use a leading order calculation for the 
Standard Model contribution $\sigma^{SM}$ in Eq.~\ref{eq:param1} 
corresponds to the Born-level
result. The charge asymmetry appears only at next-to-leading order 
in the SM, so the leading-order asymmetry in Eq.~\ref{eq:param2} is
vanishes, $A_{C}^{SM,Born} = $ 0. 
For the comparison with data NNLO+NNLL predictions are used for
the purely Standard Model contribution, while 
the charge asymmetry $A_C$ at the LHC is only available to NLO precision.} using the linear dependence of Equations~\ref{eq:param1}: 

\begin{flalign}
\begin{split}  
& \frac{(\sigma - \sigma^{SM})}{\sigma^{SM}} =   
 \\ &  \hskip 0.5cm [\alpha_u ( C_u^1 + C_u^2 ) + \alpha_d ( C_d^1 + C_d^2 ) ] (\frac{1\tev{}}{\Lambda})^2,  \label{eq:param1}
\end{split}
\end{flalign}
and~\ref{eq:param2}:
\begin{flalign}
\begin{split}  
& ( A_{C} - A_{C}^{SM} ) =   \\ & \hskip 0.5cm [\beta_u ( C_u^1 - C_u^2 ) + \beta_d ( C_d^1 - C_d^2 ) ] (\frac{1\tev{}}{\Lambda})^2 \label{eq:param2}
\end{split}
\end{flalign}

\begin{figure} 
\centering
    \subfigure[]{\includegraphics[width=0.45\textwidth]{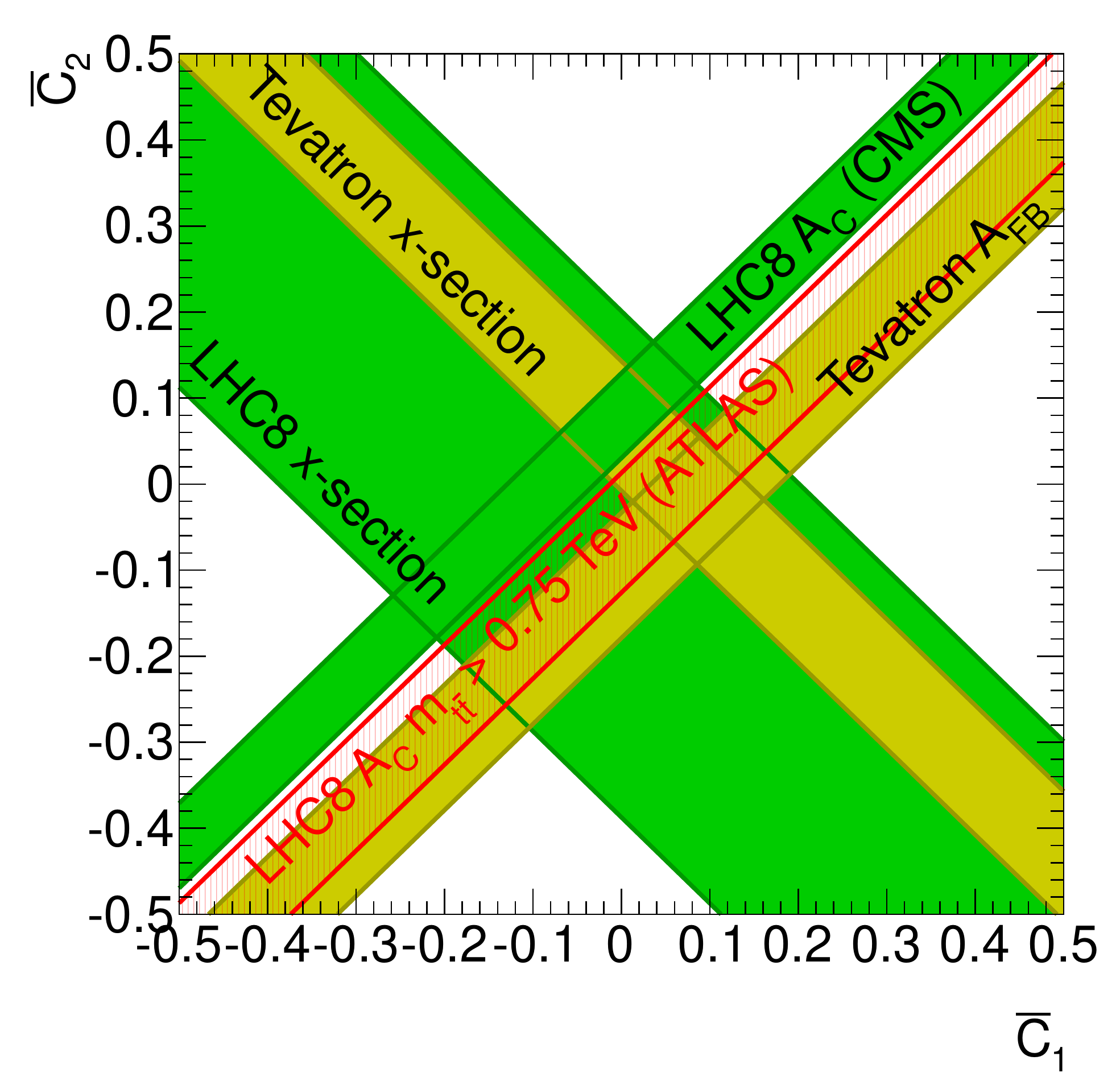}}
    \subfigure[]{\includegraphics[width=0.45\textwidth]{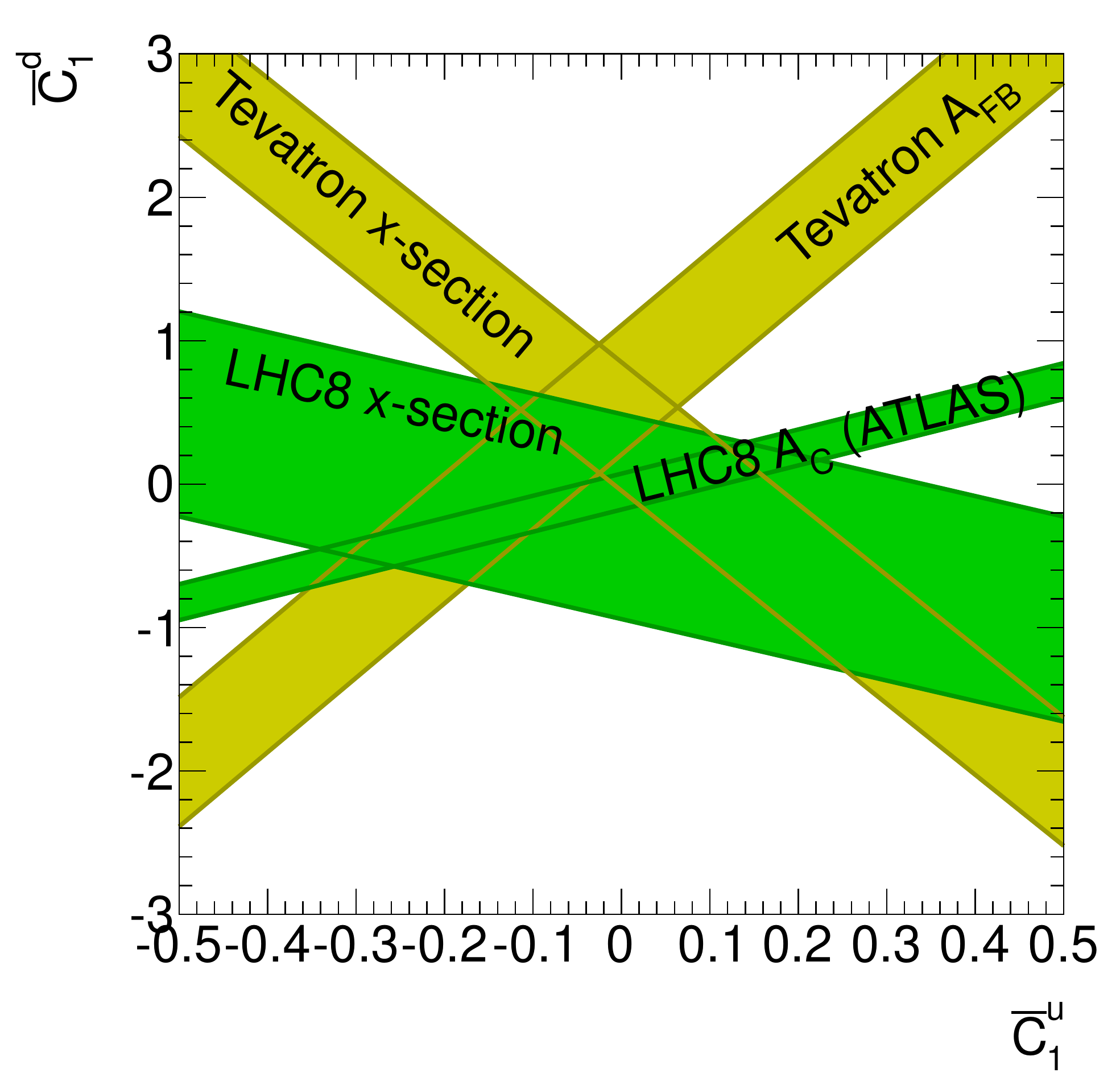}}
\caption{The constraints on pairs of effective operators from several cross-section and charge asymmetry measurements. The bands represent in (a) represent the constraints on $C^1$ and $C^2$ (assuming $C_u^1 = C_d^1 = C^1$ and $C_u^2 = C_d^2 = C^2$). The bands in (b) represent the constraints on $C_u^1$ and $C_d^1$.}
\label{fig:bands}
\end{figure}

Equation~\ref{eq:param1} shows that the cross section is proportional to 
$C^1 + C^2$, while the asymmetry in Equation~\ref{eq:param2} is proportional to $C^1 - C^2$. Therefore,
the combination of the two measurements provides a very powerful constraint
on both $C^1$ and $C^2$ operators. The complementarity is illustrated in 
Figure~\ref{fig:bands} (a), where the bands representing the constraint
from the asymmetry measurement cross the cross-section bands at a straight
angle.

The results for the coefficients of Eq.~\ref{eq:param1} and~\ref{eq:param2} are presented in 
Table~\ref{t:param}. The coefficients $\alpha_u$ and $\alpha_d$ are defined
such that they are proportional to the contribution of 
new interactions to the cross-section divided by the SM cross section. 
As such the size of $\alpha_{u/d}$ in different measurements offers a 
good indication of the sensitivity of the measurements (assuming 
the {\em relative} precision of all measurements is equal, condition 
that is approximately met for the measurements in the Table). 
The $\beta_{u/d}$ coefficients indicate the strength of the constraint 
for charge asymmetry measurements of the same absolute precision.

For all measurements the coefficients $\alpha_u$ and $\beta_u$ 
for the u-type operators are larger than $\alpha_d$ and $\beta_d$, 
that apply to d-type operators. The ratios $\alpha_u/\alpha_d$ and
$\beta_u/\beta_d$
are largest at the Tevatron, where a naive estimate 
based on the valence quark content of the proton and anti-proton 
would yield a factor of four. The large ratio at the Tevatron
is quite powerful to derive simultaneous constraints on u-type and 
d-type operators. The Tevatron bands in $C_u^1$ and $C_d^1$ space in 
Figure~\ref{fig:bands} (b) cross at more favourable angles than the LHC bands.
At the LHC (where the naive estimate 
would yield a ratio of two) the u-type and d-type operator
coefficients are much closer. 

Among the inclusive measurements, the Tevatron clearly offers a much 
greater sensitivity to four-fermion operators than the LHC at 8~\tev{},
reflecting the much larger dilution by gluon-initiated processes at the LHC.
The impact of the dilution is most clearly observed in the cross-section
bands in Figure~\ref{fig:bands} (a). Even if ATLAS and CMS have managed
to reduce the uncertainty on the pair production cross-section measurement
to approximately 4\%, the constraint from the LHC 8~\tev{} cross-section
data is quite weak.
The gluon-gluon contribution to the cross-section reaches nearly 
90\% at 13~\tev{}, reducing the sensitivity even further.

\begin{table*}[t]
\centering
\begin{tabular}{lcccc}
\hline
 & $\alpha_u$ $[\%]$&  $\alpha_d$ $[\%]$& $\beta_u$ $[\%]$& $\beta_d$ $[\%]$\\ \hline
Tevatron 1.96~\tev{} $p\bar{p}$ inclusive & $5.19\pm 0.02$ & $1.05 \pm 0.02$&$1.66 \pm 0.09$ &$0.32 \pm 0.09$\\ 
LHC 8~\tev{} $pp$ inclusive  & $1.02 \pm 0.02$ &$0.71 \pm 0.02$ & $0.37 \pm 0.09$ &$0.24 \pm 0.09$ \\ 
LHC 8~\tev{} $pp$  ($m_{\ttbar} >$ 0.75~\tev) &$3.03 \pm 0.09$& $1.56 \pm 0.09$& $2.16 \pm 0.09$ &$0.6 \pm 0.09$\\ 
LHC 13~\tev{} inclusive &  $0.73 \pm 0.02$ & $0.48 \pm 0.02$&$0.26 \pm 0.09$ &$0.27 \pm 0.09$\\ 
LHC 13~\tev{} ($m_{\ttbar} >$ 1.2~\tev) & $6.61 \pm  0.02$ & $3.60 \pm 0.02 $&$6.09 \pm  0.09$ &$2.78 \pm  0.09$\\ \hline
\end{tabular}
  \caption{Parameterization of the coefficients of Eq.~\ref{eq:param1} and~\ref{eq:param2}. The $\alpha$ and $\beta$ coefficients govern the impact of non-zero effective operators on the cross-section and the charge asymmetry, respectively. The $u/d$ subscripts indicate whether the coefficients correspond to $u-$type or $d-$type quarks.}
  \label{t:param}
\end{table*}

Table~\ref{t:param} suggests a way to restore the sensitivity of the LHC
to the level of the Tevatron and beyond. The differential measurements listed
in the table correspond to the cross-section and charge asymmetry
for boosted top quark production. For 8~\tev{} operation the phase space 
is limited to events with an invariant mass of the \ttbar{} system 
$m_{t\bar{t}} >$ 750~\gev{}. For 13~\tev{} the cut on $m_{t\bar{t}}$ is raised
to 1.2~\tev. We see that the $\alpha$ and $\beta$ coefficients of
these differential measurements are indeed an order of magnitude 
larger than those of the inclusive measurements at the same center-of-mass
energy. Therefore, the measurement of the charge
asymmetry at high mass can provide a competitive constraint, even with an 
uncertainty that is an order of magnitude larger than that of the
inclusive charge asymmetry measurement. The analysis~\cite{Aad:2015lgx} we have 
taken as an example is still statistically limited, with a non-negligible 
contribution from modelling uncertainties in these relatively unexplored
corners of phase space. With the large \ttbar{} samples that 
become available in run 2 of the LHC there is considerable margin 
for improvement of this and other differential measurements.

\section{Validity of the effective operator approach}
\label{sec:val}

The charge asymmetry is reported by several authors (see for instance 
Ref.~\cite{Bernreuther:2015yna}) to receive relatively large contributions 
from terms that are proportional to $\Lambda^{-4}$. As a full
treatment of all these terms (including the contribution of the interference between dimension-8 
operators with the SM and the interference between two dimension-6 operators vertices and the 
SM) is not feasible at present, this poor convergence may
jeopardize the effective operator paradigm in this area. 
In this Section we estimate the size of the $\Lambda^{-4}$ contributions
by calculating the contribution of the dimension-6 operator squared 
(i.e. $|BSM|^2$), which is
accessible in the TopEffTh model. We then have:
\begin{equation}
\left(O_i - O_i^{SM}\right) = A C_i \left(\frac{1\tev{}}{\Lambda}\right)^2 + A' C_i^2 \left(\frac{1\tev{}}{\Lambda}\right)^4.
\label{eq:param_sq}
\end{equation}
For each measurement and each operator from Table~\ref{t:4fo}
we determine the ratio $A/A'$. The results we obtain for the different operators 
in Table~\ref{t:4fo} are generally in good agreement for a given measurement,
but vary from one measurement to the next. We therefore present a unique interval for 
each measurement. Following Ref.~\cite{Bernreuther:2015yna} the region of validity
is given by the interval of the coefficient $C_i$ where the $\Lambda^{-2}$ 
linear term is at least twice as large as the quadratic $\Lambda^{-4}$ 
term (i.e. $A/A'> 2 C_i \left(\frac{1\tev{}}{\Lambda}\right)^2$). 

\begin{figure} 
\centering
\includegraphics[width=\columnwidth]{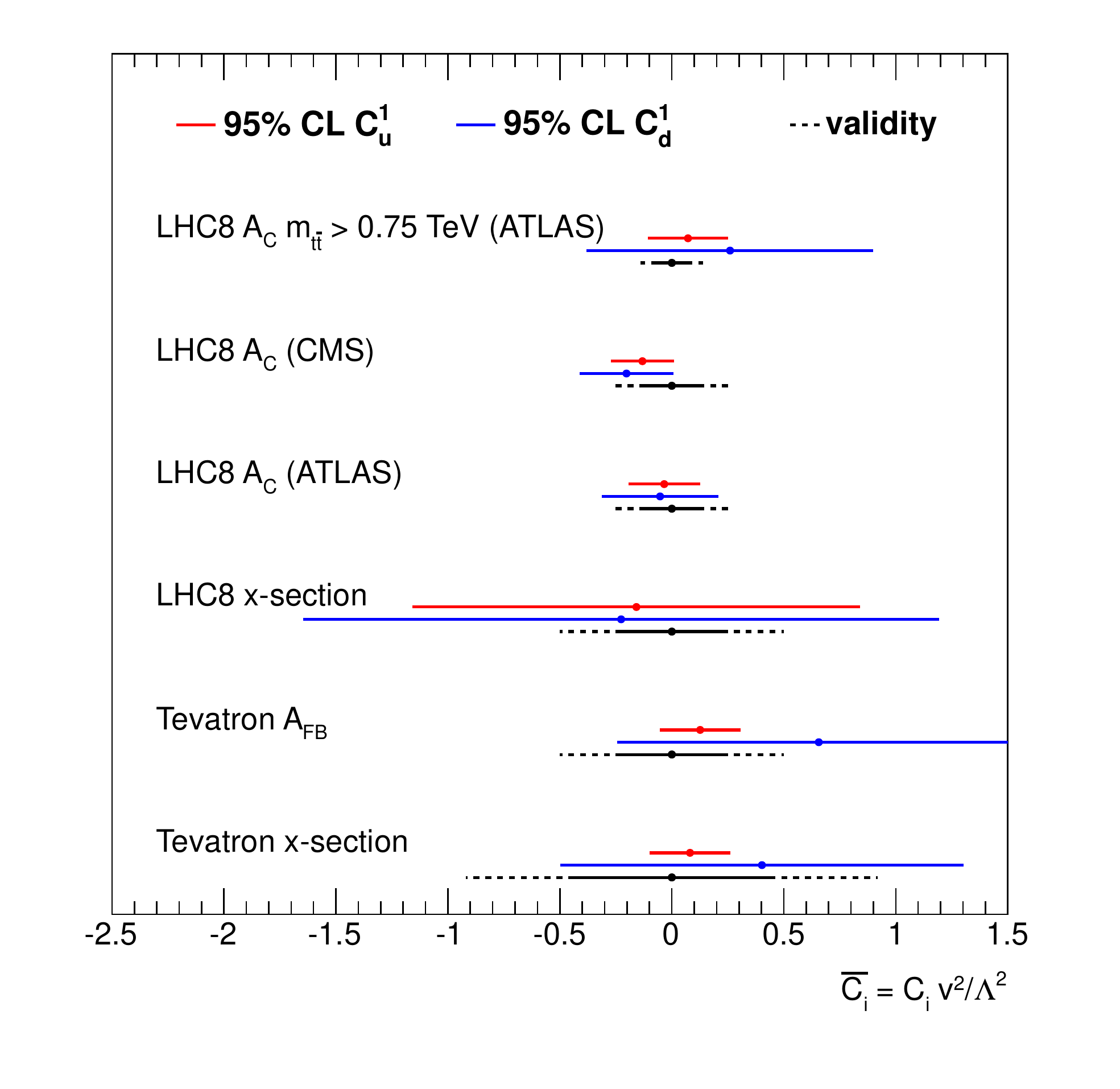}
\caption{The interval of validity and 95\% C.L. limits of cross-section and charge asymmetry measurements at hadron colliders. The interval of validity is given as a black dashed line. For each measurement the 95\% C.L. limits on the coefficients of the effective operators involving u-type and d-type quarks are indicated as error bars. Operators are fit one by one, with all other non-SM operators are set to 0. }
\label{fig:validity}
\end{figure}

In Fig.~\ref{fig:validity} the range of validity 
for each measurement is compared to the 95\% C.L. 
constraint on $C^1_u$ and $C^1_d$ 
derived from that measurement (assuming vanishing
contributions from all other operators). To guarantee valid results
we require that the 95\% C.L. interval is fully contained in the
$A/A'>$ 1 $C_i \left(\frac{1\tev{}}{\Lambda}\right)^2$ band~\footnote{This requirement ensures that the $\chi^2$ 
evaluation on the 68\% C.L. interval is within the $A/A'>$ 2 $C_i \left(\frac{1\tev{}}{\Lambda}\right)^2$ interval
where the $\Lambda^{-4}$ is of minor importance. This is therefore 
equivalent to the criterion of Ref.~\cite{Bernreuther:2015yna}.}.

The interval of validity shrinks with the increase in center-of-mass energy:
at the 8~\tev{} LHC it is typically a factor two smaller than at the Tevatron.
In combination with the reduced sensitivity to four-fermion operators
of the LHC data there is a risk that adding a measurement may reduce
the interval of validity more than the 95\% C.L. interval.
The differential measurements can see a very strong reduction of the 
interval of validity, in particular once we enter the regime of 
boosted top quark pair production. However, in this case the sensitivity 
grows to compensate the reduced interval of validity. Therefore, a
differential measurement of sufficient precision may prove to be
useful in the fit.

The same trend towards smaller interval of validity for increasing center-of-mass 
energy is observed for the cross section and the charge asymmetry.
The interval of validity of the charge asymmetry is generally somewhat smaller than
that of the cross section, but the difference is small compared to that
between the Tevatron and the LHC, or between inclusive and differential measurements. 
For the inclusive measurements at the 8~\tev{} LHC the tension between 
interval of validity and the 95\% C.L. interval on individual coefficients is 
much more pronounced for the cross-section measurement than for the
charge asymmetry.

\section{Multi-parameter fit}

So far we have evaluated constraints on one coefficient at the time, assuming
all others have a vanishing contribution. In this Section we generalize
the fit to all four-fermion operators (but still keep the remaining effective
operators related to two-fermion interactions equal to 0).
Using the parameterization, and the datasets from Table \ref{t:datasets}, 
we construct an overall $\chi^2$ function:

\begin{equation}
\chi^2 = \sum_i \left(\frac{O_i \left( \{C_i\}\right) - O_{i}^{exp}}{\Delta_{i}^{exp}}\right)^2,
\end{equation}
where $O_i \left( \{C_i\}\right)$ correponds to the parameterisation 
of Eq.~\ref{eq:param1} or Eq.~\ref{eq:param2} and $O_{i}^{exp}$ and $\Delta_{i}^{exp}$ to the 
difference between the measurement and the SM prediction. 
The sum runs over all measurements $i$ defined
in Table \ref{t:datasets}.

We minimize the $\chi^2$ function using the root package MINUIT \cite{minuit} 
in order to extract the parameters $\{C_i\}$.



The simultaneous fit of the four effective operators $\bar{C}_u^1$, $\bar{C}_u^2$, $\bar{C}_d^1$ and $\bar{C}_d^2$ 
using all data in Table~\ref{t:datasets} 
yields tight constraints on the former two, that correspond to 
interactions initiated by u-type quarks. The 95\% C.L. limits are contained
within the interval of validity.  
As we anticipated in Section~\ref{sec:sensitivity} the constraint 
on operators corresponding to d-type quarks is much weaker, where the
the marginalized 95\% C.L. constraints from the four-parameter fit 
on $\bar{C}_d^1$ and $\bar{C}_d^2$ are 3-5 times weaker than the 
limits on single operators. The marginalized 95\% C.L. intervals 
extend beyond the interval of validity. The exact level of tension
between range of validity and limits depends somewhat on which measurements
are included in the fit, but the qualitative conclusion remains true
for all combinations of the data in Table~\ref{t:datasets}: None of the 
combinations of the cross-section and charge asymmetry data yields meaningful
marginalized limits on $\bar{C}_d^1$ and $\bar{C}_d^2$. A similar observation
was made in Ref.~\cite{Buckley:2015nca} for $\bar{C}_d^2$.

Much stronger constraints are obtained when we assume $C_u^1=C_d^1=C^1$ 
and $C_u^2=C_d^2=C^2$. In this case, the interval is within the tightest 
interval of validity of the measurements used in the fit. We therefore
present the constraints obtained with the two-parameter fit as the main
result of this study.

\section{Constraints on four-fermion operators}

\begin{figure} 
\centering
\includegraphics[width=\columnwidth]{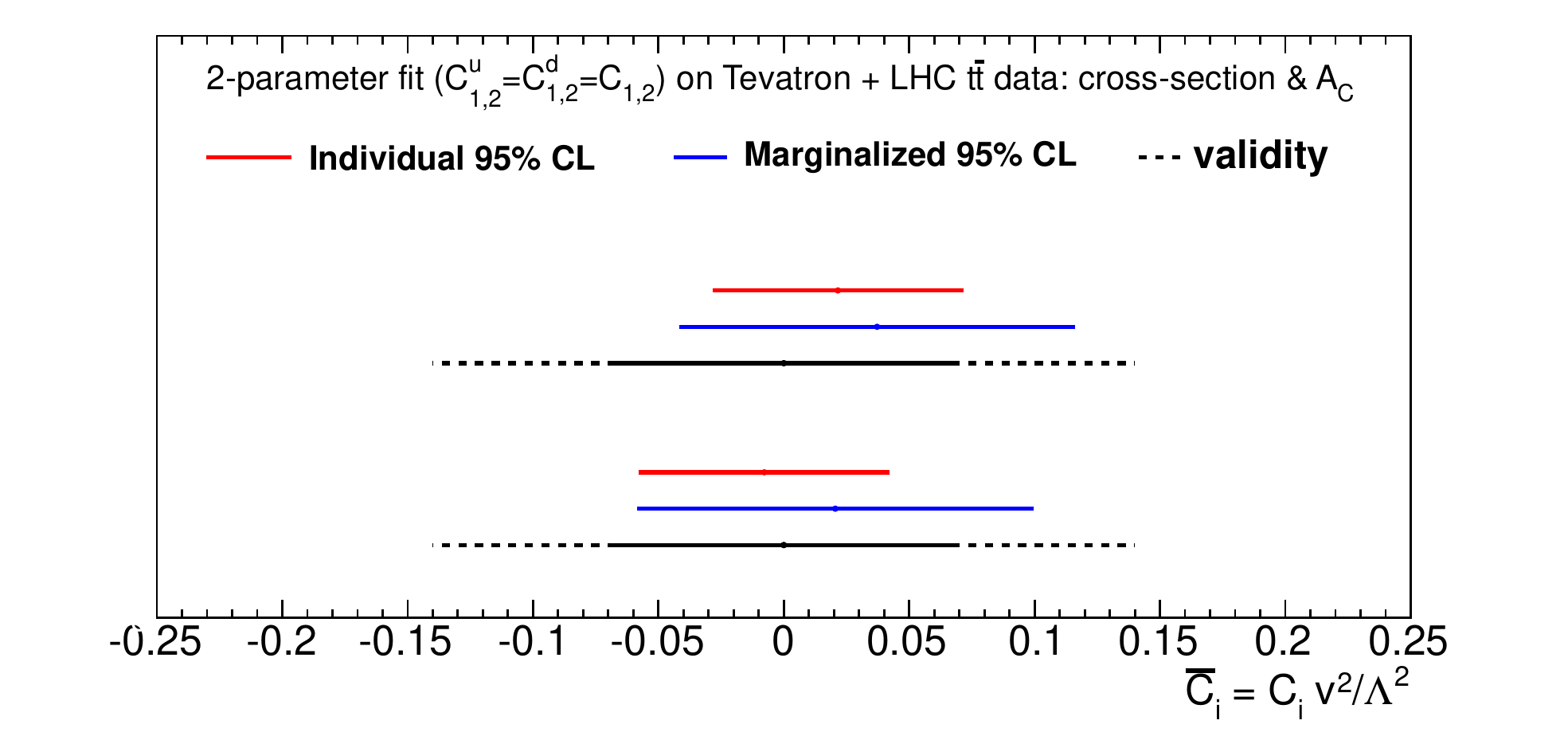}
\caption{The 95\% C.L. limits on the four-fermion operators $C_1$ and $C_2$ extracted from cross-section and charge asymmetry measurements at hadron colliders. The {\em individual} limits are obtained assuming all other non-SM operators are 0, while the {\em marginalized} limits are obtained from a two-parameter fit that floats both operator coefficients simultaneously. }
\label{fig:result}
\end{figure}

The result of the two-parameter fit of the coefficients $C^1 = C_u^1=C_d^1$ and 
$C^2=C_u^2=C_d^2$ of the four-fermion operators to \ttbar{} production cross
section and charge asymmetry measurments at hadron colliders
is presented in Figure~\ref{fig:result}. All other dimension-6 
effective operators are assumed to have negligible impact. 
The allowed intervals at 95\% confidence level are -0.06 $< \bar{C}_1 <$ 0.1 and
-0.04 $< \bar{C}_2 <$ 0.11, where $\bar{C_i} = C_i \times v^2/\Lambda^2$, with
$v=$ 246~\gev{} the Higgs vacuum expectation value.
The allowed intervals are contained within the region where 
the $\Lambda^{-2}$ contribution 
of the dimension-6 operators dominates over an estimate of the $\Lambda^{-4}$
contribution (indicated as a black line labelled ``validity'').

The allowed bands in the $C^1$-$C^2$ plane of charge asymmetry and 
cross-section measurements cross at a straight angle, yielding
tight constraints on both parameters. Indeed, the simultaneous fit 
of $C^1$ and $C^2$ yields very similar results to the limits obtained when a 
single operator is floated in the fit. 

A fit of the two linear combinations $\bar{C}_{+} = \bar{C}_{1} + \bar{C}_{2}$ 
and $\bar{C}_{-} = \bar{C}_{1} - \bar{C}_{2}$ yields limits 
-0.09 $< \bar{C}_+ <$ 0.2 and -0.07 $< \bar{C}_- <$ 0.04. In this case
the results are readily related to the measurements. We see that the
$\bar{C}_{-}$ constraint, driven by the charge asymmetry, is nearly
three times stronger than the constraint on $\bar{C}_+$, 
that is dominated by the cross-section measurements. The central value
of ${C}_+$ is 0.06, due to the Tevatron cross-section of 
Refs.~\cite{Abazov:2014cca, Aaltonen:2012it} that slightly 
exceeds the SM prediction. The ${C}_-$ fit is pulled towards negative
values by the CMS measurement in Ref.~\cite{Khachatryan:2015mna}.
This measurement 2$\sigma$ below the SM value is able to compensate the 
positive pull from the Tevatron experiments.
We propose the constraint on $\bar{C}_{-} = \bar{C}_{1} - \bar{C}_{2}$
as a benchmark for experimental analyses: the extent of the 95\% C.L. 
allowed region is a good figure-of-merit to relate the sensitivity
to high-scale new physics of measurements with different initial
states (i.e. Tevatron vs. LHC), different center-of-mass energies
and in different kinematic regimes.

The limits on the four-fermion operators presented in this paper are 
stronger than those of the global fit to the top sector presented 
in Ref.~\cite{Buckley:2015nca,Buckley:2015lku}. The prize to pay for this gain
in precision is a loss of generality: the limits we derive are valid 
only under the assumption of equal coefficients for the four-fermion
operators involving u-type and d-type quarks:
$C^1 = C_u^1=C_d^1$ and $C^2 = C_u^2=C_d^2$.
We believe, however, that this may be the most
practical way to guarantee the validity of the effective operator
approach with the current data sets. In the long run more precise 
data from LHC run 2 
should allow to constrain the separate four-fermion operators of up-type 
and down-type quarks to safe intervals\footnote{Ideally, one would float all
four degrees of freedom in the fit when extracting the coefficients of 
the two-fermion operators, so as to avoid an artificial reduction 
of the uncertainty on these parameters. Then, the four-fermion operator
constraints can be obtained under the assumption $C^1 = C_u^1=C_d^1$ and $C^2 = C_u^2=C_d^2$.}.

\section{Comparison to a concrete new physics model}

The limits on $C_-$ can be recast into limits on the mass
of a flavour-universal axigluon~\cite{Antunano:2007da} 
(with equal couplings to all quarks) using the relation 
$(C^1 - C^2)/{ \Lambda^2} = -4 g_s^2/m_{A}^2$ from
Ref.~\cite{Degrande:2010kt}. The 95\% C.L. lower limit on the 
axigluon mass is 2.0~\tev{}.
The axigluon with opposite-sign couplings to light and top quarks
($g_t = - g_q$), that makes a positive contribution to the
charge asymmetry, is even more strongly constrained: $ m > $ 2.8~\tev{}.
These limits extend the exclusion of earlier studies~\cite{Ferrario:2009bz}
considerably.

Both limites are well in excess of the 1.5~\tev{} that 
Ref.~\cite{Degrande:2010kt} as the lower limit for application
of the effective-operator analysis.

With LHC run I the sensitivity for observation of a narrow signal on the
SM \ttbar{} background has entered the sub-pb regime for a 
multi-\tev{} resonance. The ATLAS and CMS 
searches~\cite{Aad:2015fna,Khachatryan:2015sma}
yield a 95\% C.L. lower limit on the axigluon mass of order 2~\tev.
Limits from di-jet resonance searches at 13~\tev{} provide
even stronger limits on this particular model~\cite{Khachatryan:2015dcf}.
 
\section{Outlook to LHC run 2}

During the preparation of this paper the analysis of LHC run 2 data is 
in full swing. CMS has put out a first \ttbar{} cross-section 
measurement~\cite{Khachatryan:2015uqb}, 
while ATLAS has produced a preliminary result~\cite{ATLAS:2015-CONF-033}.
It is instructive to consider the effect of the inclusion of the 13~\tev{}
results in the fit.

\begin{figure} 
\centering
\includegraphics[width=\columnwidth]{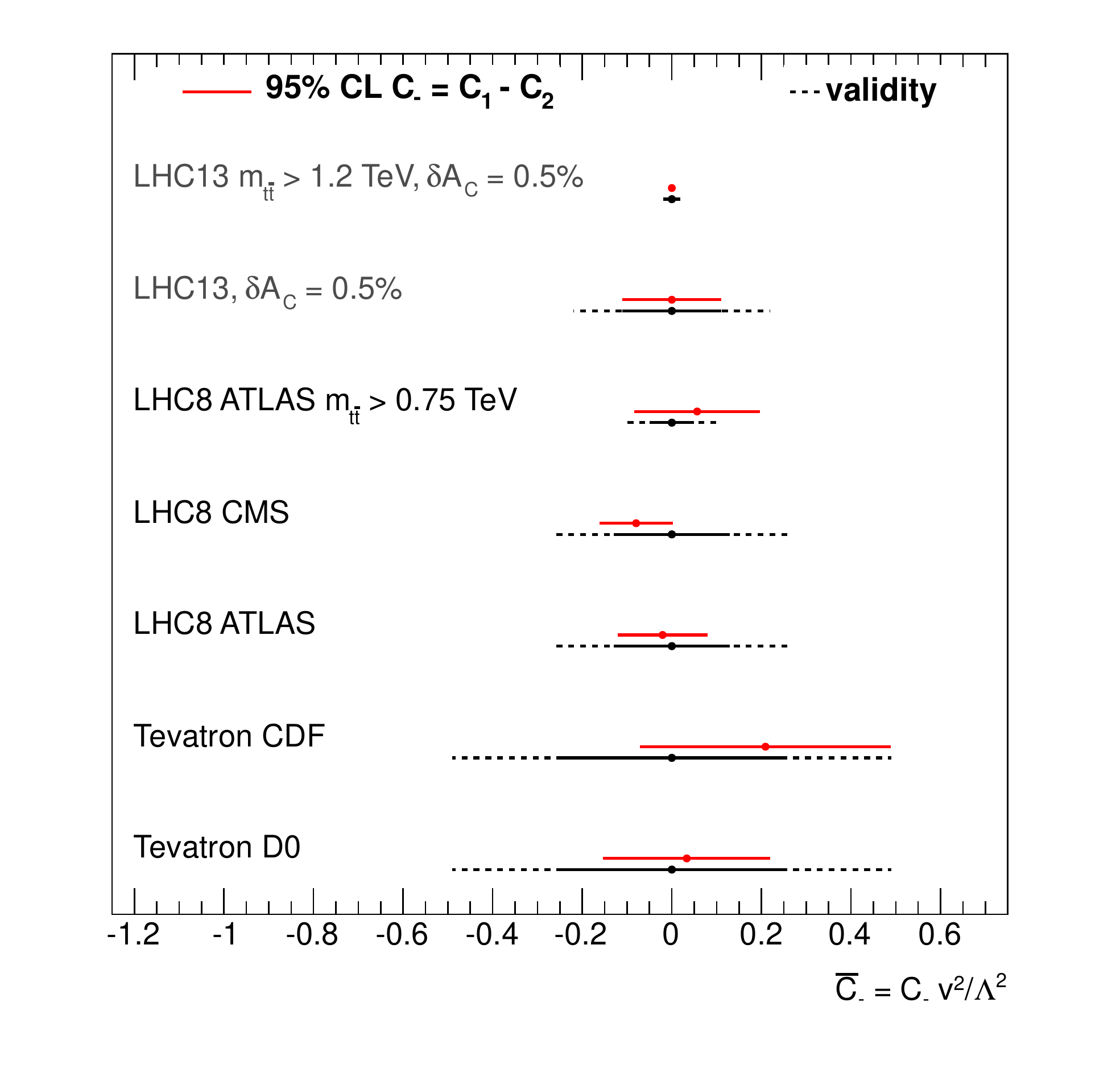}
\caption{The 95\% C.L. limits on the linear combination of four-fermion operators $\bar{C}_{-} = \bar{C}_1 - \bar{C}_2$ extracted from charge asymmetry measurements at hadron colliders. The entries labelled as LHC13 present the prospects of a charge asymmetry measurement with an uncertainty of 0.5\% and a central value in exact agreement with the SM prediction. }
\label{fig:outlook}
\end{figure}

The constraints on $\bar{C}_{-}$ of past and present measurements and the
prospects for future measurements of charge asymmetries are shown in 
Figure~\ref{fig:outlook}. 

In 13~\tev{} $pp$ collisions the $q\bar{q} \rightarrow t\bar{t}$ process
is further diluted by the increase in gluon-gluon-initiated \ttbar{} 
production. Therefore, the sensitivity of inclusive measurements to 
four-fermion operators is limited. In Figure~\ref{fig:outlook} the
expected uncertainty on $\bar{C}_{-}$ from the 13~\tev{} inclusive
charge asymmetry measurement with a precision of 0.5\% 
is larger than that of the current LHC8 measurements with a similar
precision.
With the current uncertainty of approximately 15\% 
(dominated by the 10\% uncertainty 
of the preliminary estimate of the integrated luminosity) the
cross-section measurements add no value to the fit.
For inclusive measurements the interval of validity at 13~\tev{} 
is reduced only slightly, to -0.22 $< \bar{C}_X <$ 0.22, and a 
two-parameter fit (with the assumption $C_u = C_d $) on measurements 
of comparable precision to those at 8~\tev{} is expected to 
yield a limit that remains within the interval of validity.

We already signalled in Section~\ref{sec:sensitivity} that the excellent 
sensivity to four-fermion operators of differential 
measurements, in particular measurements
in the regime of boosted \ttbar{} pair production, compensates 
for their (current) relatively poor precision. As an example, 
consider highly boosted top quark pair production with $m_{\ttbar} >$ 1.2~\tev,
the top entry in Fig.~\ref{fig:outlook}. If a charge asymmetry
is performed to 0.5\% precision an extremely tight constraint
on four-fermion interactions can be derived.  
A problem for the inclusion of such measurements is the limited
range of validity of the effective operator analysis for such measurements
(due to large contributions from the $\Lambda^{-4}$ that are 
only partially known). Requiring that
the $\Lambda^{-2}$ term dominates over $\Lambda^{-4}$ term reduces the interval 
accessible to the effective operator analysis to $|\bar{C}_X| < $ 0.03, 
well below the current limits. To constrain the measurement 
the measurement of both $\bar{C}_1$ and $\bar{C}_2$ operators
to this level would require a (relative) cross section measurement in the 
boosted regime with a precision of 4\% and a charge asymmetry measurement
with a precision of 0.5\%, which is definitely challenging, but may not
be impossible.


\section{Summary}

Top quark pair production data at hadron colliders allow to constrain
four-fermion interactions. Analyzing the relative sensitivities of
pair production measurements at the Tevatron and the LHC we find 
that the cross-section and charge asymmetry measurements 
provide complementary constraints, where
the latter are more powerful at the LHC. The sensitivity
to four-fermion operators is strongly enhanced for measurements
in the boosted regime. 

Several authors~\cite{Degrande:2010kt,Buckley:2015lku,Bernreuther:2015yna} 
have signalled the importance of higher-dimension contributions of order
${\Lambda}^{-4}$ to high-energy collision data.
We have ensured explicitly that these contributions, 
whose size is estimated as the contribution of the
dimension-6 operator squared, are subdominant in our fit. 

We have extracted limits on the dimension-6 
operators ${C}^1$ and ${C}^2$, under the assumption of that the coupling
strengths to up- and down-type quarks are identical 
(i.e. ${C}^1 = {C}^1_u = {C}^1_d$ and ${C}^2 = {C}^2_u = {C}^2_d$).
The allowed intervals at 95\% C.L., -0.06 
$ < {C}^1 \times v^2/\Lambda^2< $ 0.10 and 
-0.04 $ < {C}^2 \times v^2/\Lambda^2 < $ 0.11, are
in good agreement with the SM prediction ${C}^1 = C^2 =$ 0.
These form stricter limits than those obtained from a global fit that
includes the same data~\cite{Buckley:2015lku} (at what we
believe is an accepatable loss of generality).

For an explicit UV completion such as the axigluon model these
limits correspond to a lower limit on the mass in excess of 2~\tev, which is
a competitive constraint when compared to 
direct limits from resonance searches.

\section*{Acknowledgements}

The authors gratefully acknowledge the help of Juan Antonio~Aguilar and German~Rodrigo in the preparation of the letter, the support of Celine Degrande while setting up the TopEffTh model and helpful discussion with the members of the TopFitter collaboration.



\bibliographystyle{JHEP}
\addcontentsline{toc}{section}{References}
\bibliography{oper}

\end{document}